%% file: main.tex
\pdfoutput=1
\documentclass[runningheads, envcountsame, a4paper]{llncs}
\usepackage{enumitem}
\usepackage{mathtools}
\usepackage{amsmath}
\DeclareMathOperator*{\argmin}{argmin} 
\usepackage{subcaption}

\usepackage{booktabs} 
\usepackage{multirow} 
\usepackage{pgfplots}
\pgfplotsset{compat=1.14}

\usepackage[colorlinks]{hyperref}

\usepackage{graphicx}
\usepackage[misc]{ifsym}
\begin{document}

\title{Mend the Learning Approach, Not the Data: Insights for Ranking E-Commerce Products}

\titlerunning{Mend the Learning Approach}
\toctitle {Mend the Learning Approach for Ranking E-Commerce Products}

\author{
Muhammad Umer Anwaar (\Letter) \inst{1,3}
\and
Dmytro Rybalko\inst{2}\and
Martin Kleinsteuber\inst{1,3}}
\institute{Technische Universit{\"a}t M{\"u}nchen, Germany \and
IBM, Russia\and
Mercateo, Germany\\
\email{
umer.anwaar@tum.de, dmitriy.rybalko@ibm.com,
martin.kleinsteuber@mercateo.com
}}

\authorrunning{Anwaar et al.}

\tocauthor{Muhammad Umer~Anwaar[TUM], Dmytro~Rybalko[IBM], Martin~Kleinsteuber[Mercateo]}

\maketitle              
\setcounter{footnote}{0}
\begin{abstract}
Improved search quality enhances users' satisfaction, which directly impacts sales growth of an E-Commerce (E-Com) platform. 
Traditional Learning to Rank (LTR) algorithms require relevance judgments on products. In E-Com, getting such judgments poses an immense challenge. In the literature, it is proposed to employ user feedback (such as clicks, add-to-basket (AtB) clicks and orders) to generate relevance judgments. It is done in two steps: first, query-product pair data are aggregated from the logs and then order rate etc are calculated for each pair in the logs. In this paper, we advocate counterfactual risk minimization (CRM) approach which circumvents the need of relevance judgements, data aggregation and is better suited for learning from logged data, i.e. contextual bandit feedback. Due to unavailability of public E-Com LTR dataset, we provide \textit{Mercateo dataset} from our platform. It contains more than 10 million AtB click logs and 1 million order logs from a catalogue of about 3.5 million products associated with 3060 queries. 
To the best of our knowledge, this is the first work which examines effectiveness of CRM approach in learning ranking model from real-world logged data. Our empirical evaluation shows that our CRM approach learns effectively from logged data and beats a strong baseline ranker ($\lambda$-MART) by a huge margin. Our method outperforms full-information loss (e.g. cross-entropy) on various deep neural network models. 
These findings demonstrate that by adopting CRM approach, E-Com platforms can get better product search quality compared to full-information approach.

\keywords{
Information Retrieval \and Ranking and Preference Learning \and
Learning to Rank \and E-Commerce Search \and Implicit Feedback \and
Counterfactual Risk Minimization \and Dataset \and Mining Data Logs}
\end{abstract}

\section{Introduction}

The E-Com industry is growing fast, with a projected global sales of 4.8 trillion USD in 2021\footnote{ \url{https://www.statista.com/statistics/379046/}}. Virtually every E-Com platform leverages
machine learning (ML) techniques to increase their users' satisfaction and optimize business value for the company. Optimal ranking of search results plays a vital role in achieving these goals. 
The successful application of traditional ML algorithms, such as $\lambda$-MART \cite{chapelle2011yahoo} and AdaRank \cite{Xu2007AdaRank}, requires hand-crafted features. This greatly reduces their applicability in commercial settings due to large amount of diverse products. 
Deep learning (DL) is a well established framework for automatically learning relevant features from raw data and has also inspired research in LTR \cite{DeepRank,Duet}. But DL needs large-scale data for training a model.
Fortunately, such data is readily available in almost every E-Com platform in the form of search, clicks, add-to-basket (AtB) clicks and orders logs. We will refer to this data as \emph{log data} in the remainder of the paper.

As log data is abundantly and cheaply available, it is promising to devise learning algorithms which can learn effective ranking models from it. In contrast to online or active LTR methods \cite{schuth2013lerot,Hu:2018:RLR:3219819.3219846}, learning from log data avoids intrusive interactions in the live E-Com platform. This is highly desirable in practice because it avoids badly affecting users' experience. 
However, how to get relevance judgments (RJs) for training (offline) supervised LTR algorithms is a significant challenge in learning from log data.
RJs for benchmark LTR datasets are performed either by experts or crowd sourcing \cite{DatasetsLETOR}. 
Several studies \cite{Crowd,KarmakerSantu:2017:ALR:3077136.3080838} have shown that crowd-sourcing is not a reliable technique for getting RJs on products of E-Com platform.
Getting RJs for millions of products from domain experts is prohibitively expensive.
This has created a gap in the application of DL research for improving E-Com search quality. 

In this paper, we aim to bridge this gap and improve product search with the practical constraints of a commercial setting.
Santu et al. 
\cite{KarmakerSantu:2017:ALR:3077136.3080838} have made an attempt to overcome this issue. 
They aggregate query-product pairs from the logs and calculate user feedback rate (e.g. order rate) for these pairs. RJs are done based on this rate (for details, see Sec.~\ref{calculate}).

Such method of getting RJs ignores the fact that log data is in the form of so-called \emph{contextual-bandit} feedback. This means that we have access to \textit{only} those feedback signals which were generated in response to a limited set of actions taken by the ranking system (logging policy). For instance, we do not know how the user would have responded to the search results if another set of products was shown. That is why traditional supervised learning approach, where information about all possible actions is assumed, is not well-suited for learning from log data.
We refer to the traditional approach as full-information (Full-Info) approach and its loss (such as cross-entropy and hinge) as Full-Info loss.
This partial information (contextual-bandit feedback) challenge requires us to devise more efficient ways of utilizing the information contained in the log data.
To address this challenge effectively, the learning problem should be reformulated.

Due to these reasons, we advocate 
employing a counterfactual risk minimization (CRM) approach \cite{JMLR:v16:swaminathan15a} and adapting the LTR algorithm to learn directly from the log data.
CRM loss requires the knowledge of the logging policy, the actions taken by the logging policy and users' feedback on these actions (e.g. AtB clicks, orders etc). All this information is contained in the log data. In simple terms, logging policy means the ranker which was used by our E-Com platform and the log data is \textit{generated} by users' interactions with its actions.
We propose that CRM approach is better suited for learning from such logged contextual-bandit feedback, as it does not require full-information about all actions and their rewards. Moreover, it also circumvents the need to aggregate the log data and generate RJs.

The contributions of this work are summarized as follows:
\begin{itemize}
    \item We construct and publish a novel LTR dataset from a real-world E-Com platform.
    \item We adapt the CRM learning approach for E-Com LTR which enables learning directly from the log data.
    \item We conduct extensive experiments on the \textit{Mercateo dataset} demonstrating the effectiveness of the CRM approach in comparison to the Full-Info approach. 
\end{itemize}

\section{Related Work}
Over the past two decades, researchers have done significant amount of work on improving ranking for web search.
They also investigated the problem of getting RJs for URLs via click logs. \cite{ClickRules2,ClickRules1} used eye-tracking studies to devise a set of preference rules for interpreting the click logs. For instance, rule “Click $>$ Skip Next” means that
if user clicks on $URL_A$ at position i and skips $URL_B$ at
position i + 1 then $URL_A$ is preferable to $URL_B$. Similarly, rule “Click $>$ Click Above” means that if both are clicked then $URL_B$ is preferred over $URL_A$. Such rules generate pairwise preference judgments between URLs. \cite{ClickRulesProb} proposed a probabilistic interpretation of the click
logs based on  “Click $>$ Skip Next” and “Click $>$ Skip Above” rules.
The problem with such judgments is that they tend to learn to reverse the list of results. A click at the lowest position is preferred over all other clicks, while a click at the top position is preferred to nothing else. 
Recently, researchers \cite{recentclick1,recentclick2} have tried to overcome these issues but their focus is on improving web search with clicks and they do not take into account the unique challenges and feedback signals of E-Com search. 

Another active research direction is to devise user models for estimating how users examine the list of results \cite{UserModels1,UserModels2,UserModels3} and utilize these models to improve the automatic generation of RJs. In web search, such models are theoretically motivated and useful, however they are formulated in such a way that E-Com search can not benefit directly from them.
Despite being quite noisy, clicks are extremely important user feedback for web search. 
But in E-Com search, we have far less noisier and more informative signals like orders, AtB clicks, revenue etc.
Moreover, these approaches have to consider user intent modeling (exploratory, informative etc) and position bias.
In order to utilize click logs, models of users' click behavior or some preference rules are necessary.
Due to these issues, we exclude click signals in this work. In the presence of more valuable signals this trade-off is beneficial.

In contrast to web search, there has been little work on learning effective ranking models for E-Com search.
Santu et.al. \cite{KarmakerSantu:2017:ALR:3077136.3080838} have done a systematic study of applying traditional LTR algorithms on E-Com search. They studied the query-attribute sparsity problem, the effects of popularity based features and the reliability of relevance judgments via crowd-sourcing. They compared traditional ranking algorithms and reported that $\lambda$-MART showed best performance. 
They did not include DNN models in their work. 

In this paper, we compare two approaches to overcome the hurdle of getting RJs for E-Com. The traditional approach of supervised learning (Full-Info) advocated by \cite{KarmakerSantu:2017:ALR:3077136.3080838} and our CRM based approach of treating the log data as contextual-bandit feedback. 
Recently, \cite{deep-learning-logged-bandit-feedback} have shown that CRM loss is able to achieve predictive accuracy comparable to cross-entropy loss on object recognition tasks in computer vision. 
They replace cross-entropy loss with a self-normalized inverse propensity (SNIPS) estimator \cite{Swaminathan:2015:SEC:2969442.2969600} that enables learning from contextual-bandit feedback. 
It is worth noting that \cite{deep-learning-logged-bandit-feedback} conducted all their experiments on simulated contextual-bandit feedback on the CIFAR10 dataset. According to authors' best knowledge, this is the first work which applies SNIPS estimator on actual logged bandit feedback and verify its effectiveness in solving a real-world problem i.e. E-Com LTR.

Another major issue which has inhibited research in LTR for E-Com is the unavailability of public E-Com LTR dataset. 
Some researchers \cite{KarmakerSantu:2017:ALR:3077136.3080838,brenner2018end,Bi2019LeverageIF} have conducted experiments on E-Com datasets, but they did not publish the dataset due to data confidentiality reasons. 
We think that such a dataset is critical in advancing the research of designing effective and robust LTR algorithms for E-Com search. 
We discuss this in detail in Sec.~\ref{whyDataset}.

\section{E-Com Dataset for LTR}\label{Dataset} 
In this section, we present the dataset constructed from our Business-to-Business (B2B) E-Com platform. This dataset contains information of queries from actual users, actions taken by the logging policy, the probability of these actions and feedback of users in response to these actions. The \textit{Mercateo dataset} is publicly available to facilitate  research on ranking products of E-Com platform. The dataset can be accessed at: \url{https://github.com/ecom-research/CRM-LTR}.

\subsection{Need of a New Dataset} \label{whyDataset}
\paragraph{Unique Feedback Signals.}
E-Com platforms tackle diverse needs and preferences of their users while maximizing their profit. In practice, this translates to certain unique problems and opportunities which are irrelevant in comparable tasks like ranking documents or URLs. 
For instance, when users interact with the E-Com platform, they generate 
multiple implicit feedback signals  e.g. clicks, AtB clicks, orders, reviews etc. 
These signals can serve as a proxy for users' satisfaction with the search results and business value for the E-Com platform. 

\paragraph{Unavailablity of public E-Com LTR datasets.}
To the best of our knowledge, there is no publicly available E-Com LTR dataset. The principal reason is that such datasets may contain confidential or proprietary information, and E-Com platforms are unwilling to take this risk.
For instance, \cite{KarmakerSantu:2017:ALR:3077136.3080838,brenner2018end,Bi2019LeverageIF} worked on LTR for E-Com search with \textit{Walmart and Amazon} datasets but they did not publish it.
There are some public E-Com datasets for problems like clustering or recommendation systems \cite{Dua:2017}, \cite{Sidana:2017:KLD:3077136.3080713}, but they have very few features (e.g. 8 features in \cite{lsbupr1492}) and few products to be considered viable for the LTR task.
\paragraph{Logged dataset.} 
Although CRM loss does not need any RJs for training, it still requires logged contextual bandit feedback. Thus, we also publish the real-world logged dataset, which will prove instrumental in advancing the research on learning directly from logs.


\begin{table}
\centering
\caption [Statistics for the \textit{Mercateo dataset}]{Statistics for the \textit{Mercateo dataset}}

\begin{tabular}{ |l|c| }
\hline
Total \# of Queries &  3060\\ \hline
\# of Queries in [Train/Dev/Test] set & [1836/612/612]\\ \hline
\# of Products in Dataset & 3507965\\ \hline 
Avg. \# of Products per Query [Train/Dev/Test] & [1106/1218/1192] \\ \hline
\end{tabular}
\label{MercateoDatasetStats1}

\end{table}

\subsection{Scope of the Dataset} \label{scope}

For a given query, E-Com search can be divided into two broad steps. The first step retrieves potentially relevant products from the product catalogs and the second step ranks these retrieved products in such a fashion which optimizes users' satisfaction and business value for the E-Com platform \cite{LTRsteps}.
Our proprietary algorithm performs the first part effectively. Thus, the scope of this dataset is limited to overcoming the challenges faced in optimizing the second part of the E-Com search.
Another important thing to consider is that most of our users (buyers) are small and medium-sized businesses (SMBs). On our platform, we employ user-specific information to personalise the final ranking based on the contracts of buyers with suppliers. Such user information is proprietary and it is against our business interests to publish it even after anonymising.  Therefore, we exclude user information in this dataset.
\paragraph{User Feedback Signals}
Although there are plenty of feedback signals logged by E-Com platforms, we considered only two common feedback signals in our dataset. One is AtB clicks on the results shown in response to a query and the second is the order signal on the same list of results. 
AtB clicks serve as implicit feedback on search quality, i.e. the user found the product interesting enough to first click it on the search result page and then click again to put it in the basket (cart). 
Orders, though more sparse than AtB clicks, are less noisy and correlate strongly with users' satisfaction with E-Com search results. They also serve as proxy to business value for the E-Com platform.
The \textit{Mercateo dataset} contains more than 10 million and 1 million AtB click and order log entries respectively. 
Table~\ref{MercateoDatasetStats1} shows some statistics about the dataset.

\subsection{Dataset Construction} \label{construction}
\paragraph{Data sources} \label{collection} 
We collected data from sources commonly available in E-Com platforms. Namely: (i) title of the products, (ii) AtB click data from logs, (iii) search logs (containing information about the products displayed on the search result page in response to the query) and (iv) order logs 

\paragraph{Selecting the queries} \label{queries}
Most queries included in this dataset are those which were challenging for our current ranking algorithm. Queries were subsampled from the search logs, keeping in mind the following considerations: (i) ensure statistical significance of the learning outcomes (especially for dev and test sets, i.e. must have been searched at least 1000 times) and (ii) include queries with variable \textit{specificity}. We say a query has low \textit{specificity}, if the query entered by the user has a broad range of products associated with it. This can be ensured simply by looking at the number of \emph{potentially relevant products} returned by our proprietary algorithm. For instance, for a very specific query like \textit{pink painting brush} (20) products are returned, while for broad and common queries like \textit{iphone} (354) or \textit{beamer} (1,282) products are returned. 


\paragraph
{How to get relevance judgments (RJs) on the products for supervised dataset?} \label{Justification} 
Our CRM based approach does not require RJs for learning but RJs are required for Full-Info loss. These RJs are also needed for a fair comparison of CRM approach with Full-Info approach. But how do we get them for millions of products?
RJs for benchmark datasets \cite{DatasetsLETOR,chapelle2011yahoo} are performed by human judges; who can be domain experts or crowd-source workers. Unfortunately it is too expensive to ask experts to judge products in an E-Com setting.

The problems associated with getting RJs via crowd-sourcing have been analyzed extensively by \cite{Crowd} using Amazon Mechanical Turk. They found that users of E-Com platforms have a very complex utility function and their criteria of relevance may depend on product's value for money, brand, warranty etc. Crowd-source workers fail to capture all these aspects of relevance. 
Hence, crowd-sourcing is an unreliable method for getting RJs.
This was also confirmed by another study \cite{KarmakerSantu:2017:ALR:3077136.3080838}.

An alternative approach is to generate RJs from multiple feedback signals present in user interaction logs and historical sales data of products. 
Major benefits of this approach are: (i) such RJs are closer to the notion of relevance in E-Com search and quantify relevance as a proxy of user satisfaction and business value, (ii) it costs less time and money, as compared to other two approaches, and (iii) large-scale E-Com datasets can be constructed, as it only requires data generated by users' interaction with the E-Com website. Such data is abundantly available in almost every E-Com platform. 
One drawback of this approach is that the quality of these RJs may not be as good as human expert judgments.
Based on the analysis of available choices for the supervised setting, generating RJs from logs in such a manner is justified. \cite{KarmakerSantu:2017:ALR:3077136.3080838} also advocated this approach.

\paragraph{Calculating Relevance Judgments (Labels)} \label{calculate}

For a given query $q$, we now show the steps taken to calculate relevance judgments on the set of products, $\mathcal{P}_{q}$, associated with it:
\begin{enumerate}
\item We calculate the visibility of a product, i.e. the number of times a product $p$ was shown to the users. We remove the products from our dataset with visibility less than 50.
\item We then convert AtB click (order) signal to \textit{relevance rate} (RR) by dividing the AtB clicks (orders) for product $p$, with its visibility. 
\item Next, we normalize RR with the maximum value of RR for that query $q$ to get normalized relevance rate (NRR).
\end{enumerate}

We publish NRR for all query-product pairs. This allows researchers the flexibility in computing different types of relevance judgments (e.g. binary, graded, continuous etc).
We follow \cite{KarmakerSantu:2017:ALR:3077136.3080838} and compute the 5-point graded relevance judgment, $l(p,q)$, by the following formula:
\begin{equation} \label{graded}
l(p,q) =  ceil [4 \cdot {NRR(p,q)}]
\end{equation}


\paragraph{Dataset format} \label{extractfeatures} 



Although there are plenty of feedback signals logged by E-Com platforms, we considered only two common feedback signals in our dataset. One is AtB clicks on the results shown in response to a query and the second is the order signals on the same list of results. 
For each query, we retain all products that were AtB clicked (ordered) and negatively sample remaining products from the search logs, i.e. products which were shown to the user but not AtB clicked (ordered).
A supervised train set is also published along with AtB click and order labels. 
The queries and the products are the same for both the supervised train and the logged training sets. Test and development (dev) sets are published \textit{only} with supervisory labels. The split of queries among train, dev and test sets is done randomly with 60\%, 20\% and 20\% of total number of queries. There is no overlap of queries in training, dev and test sets.
Raw text can not be published because our sellers have the proprietary rights on the product title and description. We publish 100-dimensional GloVe word embeddings trained on the  corpus comprising of queries and product titles. 
For each query-product pair, we also provide some proprietary features which can be employed as \textit{dense features} in DNNs. These features contain information about price, delivery time, profit margin etc of the product. Specific details can not be shared due to confidentiality reasons. Further details can be found on the github repository.

\section{Problem Formulation} \label{CRM}
In Full-Info setting, e.g. pointwise LTR, a query-product pair and its RJ defines the training instance. The task of the ranker is to predict RJ for a given query-product pair. 
In our CRM setting, we say that a query-product pair defines the context. Additional information such as users' model for personalized ranking, product description etc, can also be incorporated into the context. But for reasons discussed in Sec.~\ref{Dataset}, we limit the context to search query, product title and dense features.
The ranking system (logging policy) can only take binary actions given the context. That is, whether the product is to be displayed among the top-$k$ positions or not. These actions incur reward (loss) based on feedback signals from the user. 
It is to be noted that the top-$k$ positions are extremely important for real-world LTR systems. In E-Com platforms, $k$ is usually not the same for all queries. It depends on the query specificity determined by our proprietary algorithm. For instance, too broad a query like \textit{copy paper} signals informational user intent. Thus, $k$ has a big value for such queries in comparison to specific queries like \textit{pink painting brush}.

Formally, let context $c \in \mathcal{C}$ be word embeddings of the search query, word embeddings of the product title and dense features. Let $a \in \mathcal{A}$ denote the action taken by the logging policy.
The probability of $a$ given the context $c$ is determined by the logging policy denoted as $\pi_0(a|c)$ running on the E-Com platform. For a sample $i$, this probability is also known as propensity $p_i$, i.e., $p_i=\pi_0(a_i|c_i)$. 
Based on the user feedback, a binary loss denoted as $\delta_i$ is incurred.
If $a=1$ is selected by $\pi_0$ for a context (query-product pair) and the user AtB clicked (ordered), it implies $\delta=0$. This is \emph{positive feedback}, i.e., decision of $\pi_0$ is correct and product is relevant to the user. 
On the other hand, if $a = 0$ and the user went beyond top-k results and AtB clicked (ordered) that product, we set $\delta = 1$. 
This is \emph{negative feedback}, i.e., the product was relevant and $\pi_0$ made a wrong decision to not show the product in top-k results. 
The logged data is a collection of $n$ tuples:
$D = [(c_1, a_1, p_1, \delta_1),...,(c_n, a_n, p_n, \delta_n)].$

The goal of counterfactual risk minimization is to learn an unbiased stochastic policy $\pi_w$ from logged data, which can be interpreted as a conditional distribution $\pi_w(A|c)$ over actions $a \in \mathcal{A}$.
This conditional distribution can be modeled by a DNN, $f_w(\cdot)$, with a softmax output layer:
\begin{equation}
\pi_w(a|c) = \frac{exp(f_w(c,a))}{\sum_{a^{'}\in \mathcal{A}} exp(f_w(c, a^{'}))}.
\end{equation}


Swaminathan et.al. \cite{Swaminathan:2015:SEC:2969442.2969600} proposed SNIPS as an efficient estimator to counterfactual risk. For details, we refer to \cite{deep-learning-logged-bandit-feedback}. We use the SNIPS estimator in all our experiments:
\begin{equation} \label{eq:snips}
    \hat{R}_{SNIPS}(\pi_w) = \frac{\sum_{i=1}^{n} \delta_i \frac{\pi_w(a_i| c_i)}{\pi_0(a_i|c_i)}}{ \sum_{i=1}^{n} \frac{\pi_w(a_i| c_i)}{\pi_0(a_i|c_i)}}.
\end{equation}

\section{Experiments} \label{Experiments}
We evaluate our proposed method on the real-world \textit{Mercateo dataset}. We aim to answer
the following research questions:
\begin{itemize}
    \item \textbf{RQ1:} How does our CRM-based method perform compared to the Full-Info approach and $\lambda$-MART? 
    \item \textbf{RQ2:} How does the performance of our method progress with more bandit feedback?
    \item \textbf{RQ3:} How does the DNN architecture affect the performance of CRM and Full-Info Loss?
\end{itemize}
\subsection{Experimental Setup}

In our experiments, we use mini-batch Adam and select the model which yields best performance on the dev set. Further implementation details can be found in the code\footnote{Available at: \url{https://github.com/ecom-research/CRM-LTR}}.
For evaluation, we use the standard \textit{trec\_eval} \footnote{Available at: \url{https://github.com/usnistgov/trec_eval}} tool and several popular evaluation metrics.

\subsection{Comparison of CRM and Full-Info Approaches (RQ1)}

In this experiment, we selected a simple yet powerful CNN model \cite{Severyn:2015:LRS:2766462.2767738}, which we refer to as S-CNN. It utilizes convolutional filters for ranking pairs of short texts. We compare the performance of CRM loss with Full-Info (cross-entropy) loss on S-CNN model. All network parameters are kept the same for fair comparison. 

We also report the performance of two strong baseline rankers, namely logging policy and $\lambda$-MART \cite{Wu2010Adapting}. 
It has been recently shown that $\lambda$-MART outperforms traditional LTR algorithms on an E-Com dataset \cite{KarmakerSantu:2017:ALR:3077136.3080838}. 
For $\lambda$-MART, we use the open source $RankLib$ toolkit\footnote{Available at: \url{https://sourceforge.net/p/lemur/wiki/RankLib/}} with default parameters. $\lambda$-MART takes dense features engineered by domain experts as input. The logging policy 
also employs the same dense features. Predictions of this policy are used by our CRM method for actions $a_i$ and $\pi_0(a_i|c_i)$.

Tables~\ref{click_results_graded} and \ref{tab:order_results} summarize the results of the performance comparison on the test set with graded AtB click label and order label respectively. 
First, we note that S-CNN performs significantly better than the
logging policy. 
In terms of MAP, the performance improvement against logging policy is 7.8\% with Full-Info (cross-entropy) loss and 27.4\%  with CRM loss for target label AtB clicks. For target label orders, it is 32.6\% with Full-Info loss and 54.6\% with CRM loss.

Second we observe that the $\lambda$-MART model performs worse on all metrics. It performs significantly worse than even logging policy. 
Particularly, performance degradation is huge on NDCG@10 in comparison to MAP. This huge difference in metrics suggest that $\lambda$-MART has failed to learn graded relevance. Since MAP treats all relevance other than zero as one, whereas NDCG metric is sensitive to graded relevance.
These results also suggest that a deep learning approach can significantly outperform $\lambda$-MART, given enough training data. This is consistent with findings of \cite{DeepRank}.
One can argue that $\lambda$-MART has access to only few hand-crafted features, so its performance can be improved by adding more features. But feature engineering requires domain expertise and is a time-consuming process. Moreover, our logging policy employs exactly the same features and is performing quite better than $\lambda$-MART.

  \begin{table*}
  \centering
     \caption{Performance comparison for target label: AtB clicks. Significant degradation with respect to our implementation (p-value $\leq$ 0.05)}

   \begin{tabular}{lrrrrrr}
     \toprule
     Ranker (Loss) & \qquad MAP & \qquad MRR & \qquad P@5 & \quad P@10 &  NDCG@5 &  NDCG@10\\
     \midrule
     Logging policy & 0.4704 & 0.6123 & 0.4686 & 0.4613 & 0.2052 & 0.2537 \\
     $\lambda$-MART & 0.3825 & 0.4472 & 0.2972 & 0.3119 & 0.0917 & 0.1164 \\ 
     S-CNN (Full-Info) & 0.5074 & 0.8036 & 0.6552 & 0.6261 & 0.3362 & 0.3835\\
     \textbf{S-CNN (CRM) - ours} & \textbf{0.5993} & \textbf{0.8391} & \textbf{0.7346} & \textbf{0.7093} & \textbf{0.4332} & \textbf{0.4964} \\ 
     \bottomrule
   \end{tabular}
\label{click_results_graded}
      
 \end{table*}
 
 Third we note that our CRM based approach outperforms Full-Info loss on all metrics by a huge margin. Concretely, on NDCG@10, the performance gain of our CRM method over Full-Info loss is 29.4\% for AtB click labels and 14.9\% for order labels.
These results show that our CRM approach not only learns effectively from logged data but also significantly outperforms the models trained in supervised fashion on aggregated data, i.e. Full-Info approach.
  \begin{table*}

    \centering
      \caption{Performance comparison for target label: orders.  Significant degradation with respect to our implementation (p-value $\leq$ 0.05)}
   \begin{tabular}{lrrrrrr}
     \toprule
     Ranker (Loss) & \qquad MAP & \qquad MRR & \qquad P@5 & \quad P@10 &  NDCG@5 &  NDCG@10\\
     \midrule
    Logging policy & 0.2057 & 0.3225 & 0.1693 & 0.1717 & 0.0945 & 0.1250 \\
    $\lambda$-MART & 0.1519 & 0.1772 & 0.0747 & 0.0884 & 0.0322 & 0.0479 \\
    S-CNN (Full-Info) & 0.2728 & 0.4601 & \textbf{0.2869} & 0.2562 & 0.1720 & 0.1973\\
    \textbf{S-CNN (CRM) - ours} & \textbf{0.3181} & \textbf{0.4609} & 0.2841 & \textbf{0.2791} & \textbf{0.1854} & \textbf{0.2266}\\ 
    \bottomrule
  \end{tabular}
    \label{tab:order_results}
          
  \end{table*}

\begin{figure}
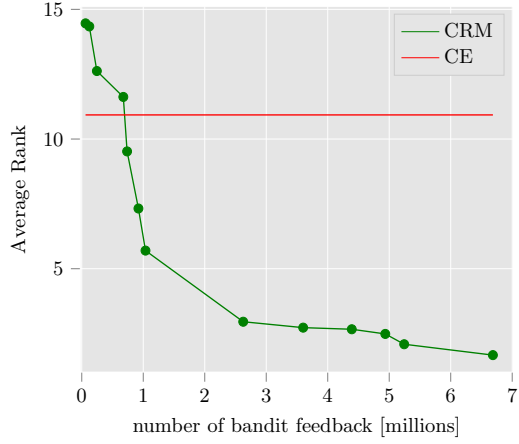

    \centering

    \scalebox{0.99}{\InputIfFileExists{plots/average_rank.tikz}{}{\textbf{!! Missing graphics !!}}}
    \caption[Average rank of relevant product in AtB click test set]{Average rank of relevant product in AtB click test set}
    \label{fig:AvgRank}

\end{figure}
\subsection{Learning Progress with Increasing Number of Bandit Feedback  (RQ2)}

It is quite insightful to measure how the performance of our CRM model changes with more bandit feedback (e.g. AtB clicks). 
To visualize this, we pause the training of S-CNN after a certain number of bandit feedback samples, evaluate the model on the test set and then resume training.

\begin{figure}
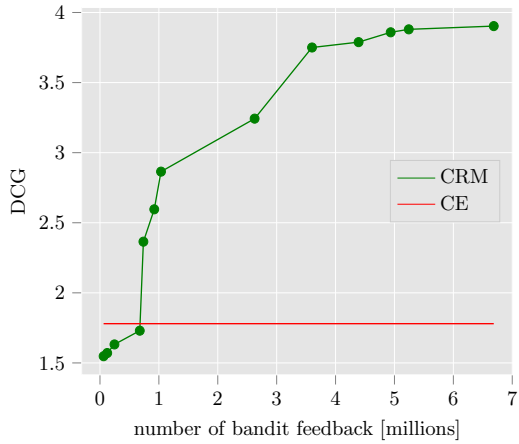

    \centering
    \scalebox{0.99}{\InputIfFileExists{plots/dcg_1.tikz}{}{\textbf{!! Missing graphics !!}}}
    \caption[Average DCG of relevant product in AtB click test set]{Average DCG of relevant product in AtB click test set}
    \label{fig:AvgDCG}

\end{figure}

In Fig.~\ref{fig:AvgRank} and Fig.~\ref{fig:AvgDCG}, each green dot corresponds to these intermediate model evaluations.
We plot average rank and average Discounted Cumulative Gain (DCG) scores of relevant products for all queries in AtB clicks test set.
These curves show that there is a constant improvement in average rank of relevant (AtB clicked) product as more bandit feedback is processed. Similarly, average DCG values rise monotonically with increasing number of AtB click feedback. Due to space constraints, we omit the plot for orders but they follow a similar pattern. The red line in these figures correspond to Full-Info S-CNN model trained on complete training set.
This experiment demonstrates the ability of the CRM approach to learn efficiently with increasing bandit feedback.

\subsection{Effect of the DNN Architecture  (RQ3)}
In order to investigate whether the improvement in performance is architecture agnostic or not, we compare CRM loss with Full-Info loss on different DNN architectures. We choose four models from MatchZoo \cite{MatchZoo:Guo:2019:MLP:3331184.3331403}, an open-source codebase for deep text matching research. For a fair comparison with S-CNN model, we modified the models in MatchZoo and added a fully connected layer before the last layer. This layer is added so that we can utilize the dense features. Results are summarized in Table~\ref{tab:DeepModelsorder_results}.

We note that the CRM loss outperfoms Full-Info loss on all of these DNN models. Specifically, performance gain of CRM loss over Full-Info loss w.r.t NDCG@10 is 10.5\% for ARCII, 13.3\% for DRMMMTKS, 7.4\% for ConvKNRM and 6.6\% for DUET. It shows that performance gains achieved by our method are not limited to any specific architecture. 

We also observe that the deep learning model architecture has significant impact on the performance.
For instance, the best performing model with Full-Info loss, DUET  has 27.8\% improvement over S-CNN w.r.t MAP. 
For CRM loss, the best performing model DUET  has 21\% improvement over the worst performing model ConvKNRM  w.r.t MAP. 

The results of this experiment support our claim that adapting 
the learning approach to learn directly from logged data is beneficial as compared to modifying the data to fit the supervised (Full-Info) learning approach.

\begin{table*}
    \centering
    \caption{Comparison of Deep Learning Models for target label: orders.}
  \begin{tabular}{ccrrrr}
    \toprule
    Ranker & Loss &  \qquad   & \qquad  MAP & \quad NDCG@5 & \quad NDCG@10\\
    \toprule
    S-CNN \cite{Severyn:2015:LRS:2766462.2767738}  & Full-Info && 0.2728 &  0.1720 & 0.1973\\
    \textbf{S-CNN} \cite{Severyn:2015:LRS:2766462.2767738} & \textbf{CRM} && \textbf{0.3181} & \textbf{0.1854} & \textbf{0.2266}\\ 
    ARCII \cite{ARCII:DBLP:journals/corr/WanLGXPC15} & Full-Info && 0.2891 & 0.2044 & 0.2238 \\
    \textbf{ARCII} \cite{ARCII:DBLP:journals/corr/WanLGXPC15} & \textbf{CRM} && \textbf{0.3208} & \textbf{0.2319 }& \textbf{0.2472} \\
    DRMMTKS \cite{DRMMTKS:10.1007/978-3-030-01012-6_2} & Full-Info && 0.3112 & 0.2183 & 0.2309 \\
    \textbf{DRMMTKS} \cite{DRMMTKS:10.1007/978-3-030-01012-6_2} & \textbf{CRM} && \textbf{0.3361} & \textbf{0.2436} & \textbf{0.2617} \\
    ConvKNRM \cite{ConVKNRM:Dai:2018:CNN:3159652.3159659} & Full-Info && 0.2818 & 0.1159 & 0.1494 \\
    \textbf{ConvKNRM} \cite{ConVKNRM:Dai:2018:CNN:3159652.3159659} & \textbf{CRM} && \textbf{0.2942} & \textbf{0.1344} & \textbf{0.1604 }\\
    DUET \cite{Duet} & Full-Info  && 0.3488 & 0.2501 & 0.2866 \\
    \textbf{DUET} \cite{Duet} & \textbf{CRM} && \textbf{0.3562} & \textbf{0.2679} & \textbf{0.3055} \\
    \bottomrule
  \end{tabular}
    \label{tab:DeepModelsorder_results}

\end{table*}

\section{Conclusion}
In E-Com platforms, user feedback signals are ubiquitous and are usually available in log files. These signals can be interpreted as contextual-bandit feedback, i.e. partial information which is limited to the actions taken by the logging policy and users' response to the actions. In order to learn effective ranking of the products from such logged data, we propose to employ counterfactual risk minimization approach. 
Our experiments on \textit{Mercateo dataset} have shown that CRM approach outperforms traditional supervised (full-information) approach on several DNN models. On \textit{Mercateo dataset}, it shows empirically that reformulating the LTR problem to utilize the information contained in log files is a better approach than artificial adapting of data for the supervised learning algorithm.
\section*{Acknowledgments}

We would like to thank Alan Schelten, Till Brychcy and Rudolf Sailer for insightful discussions which helped in improving the quality of this work.
This work has been supported by the Bavarian Ministry of Economic Affairs, Regional Development and Energy through the \emph{WoWNet} project  IUK-1902-003// IUK625/002.

\clearpage
\appendix

\section{Comparison of counterfactual risk estimators}

We compare the performance of SNIPS estimator with two baseline estimators for counterfactual risk. 
We conduct the experiments on AtB click training data of \textit{Mercateo dataset}. 
The inverse porpensity scoring (IPS) estimator is calculated by:
\begin{equation} \label{eq:ips_estimator}
\hat{R}_{IPS}(\pi_w) = \frac{1}{n}\sum_{i=1}^n \delta_i \frac{\pi_w(a_i|c_i)}{\pi_0(a_i|c_i)}.
\end{equation}
\noindent Second estimator is an empirical average (EA) estimator defined as follows:
\begin{equation}
\hat{R}_{EA}(\pi_w) = \sum_{(c,a) \in (\mathcal{C,A})} \overline{\delta} (c,a) \pi_w (a|c) ,
\end{equation}
where $\overline{\delta} (c,a)$ is the empirical average of the losses for a given context and action pair. The results for these estimators are provided in Table~\ref{tab:other_estimators}. Compared to SNIPS both IPS and EA perform significantly worse on all evaluated metrics. The results confirm the importance of equivariance of the counterfactual estimator and  show the advantages of SNIPS estimator.
 \begin{table*}
  \centering
      \caption{Results on \textit{Mercateo dataset} with AtB click relevance for IPS and empirical average estimators}
  \begin{tabular}{lcccc}
     \toprule
     Estimator \qquad \qquad &  MAP & MRR & NDCG@5 & NDCG@10\\
     \midrule
     \textbf{CNN (CRM) - SNIPS} & \textbf{0.5993} & \textbf{0.8391} & \textbf{0.4332} & \textbf{0.4964} \\ 
     CNN (CRM) - IPS & 0.4229 & 0.7139 & 0.3426 & 0.3703\\
     CNN (CRM) - EA & 0.2320 & 0.3512 &  0.2083 & 0.2253\\
     \bottomrule
  \end{tabular}
  \label{tab:other_estimators}
 \end{table*}
\begin{figure}
\centering    
\begin{minipage}{.5\textwidth}
  \centering
  \scalebox{0.75}{\InputIfFileExists{plots/snips_denominator.tikz}{}{ss}}
  \captionof{figure}{SNIPS denominator vs $\lambda$ on \\ order logs (training set)}
    \label{fig:Sandlambda}
\end{minipage}%
\begin{minipage}{.5\textwidth}
  \centering
  \scalebox{0.75}{\InputIfFileExists{plots/lagrange_map_ndcg.tikz}{}{sss}}
  \captionof{figure}{Performance on orders test set of rankers trained with different $\lambda$}
    \label{fig:Metricsandlambda}
\end{minipage}
\end{figure}

\section{Choosing hyperparameter $\lambda$}
One major drawback of SNIPS estimator is that, being a ratio estimator, it is not possible to perform its direct stochastic optimization \cite{deep-learning-logged-bandit-feedback}.
In particular, given the success of stochastic gradient descent (SGD) training of deep neural networks in related applications, this is quite disadvantageous as one can not employ SGD for training.

To overcome this limitation, Joachims et. al. \cite{deep-learning-logged-bandit-feedback} fixed the value of denominator in Eq.~\ref{eq:snips}. They denote the denominator by  $S$ and solve multiple constrained optimization problems for different values of $S$ . 
Each of these problems can be reformulated using lagrangian of the constrained optimization problem as:
\begin{equation} \label{eq:lagrange_min}
    \hat{w}_j = \argmin_{w} \frac{1}{n} \sum_{i=1}^n (\delta_i - \lambda_j) \frac{\pi_w(a_i|c_i)}{\pi_0(a_i|c_i)}
\end{equation}
where $\lambda_j$ corresponds to a fixed denominator $S_j$. 

The main difficulty in applying the CRM method to learn from logged data is the need to choose hyperparameter $\lambda$. We discuss below our heuristics of selecting it. We also evaluate the dependence of $\lambda$ on SNIPS denominator $S$, which can be used to guide the search for $\lambda$. 
To achieve good performance with CRM loss, one has to tune hyperparameter $\lambda \in [0, 1]$. Instead of doing a grid search, we follow a smarter way to find a suitable $\lambda$.
Building on the observations proposed in \cite{deep-learning-logged-bandit-feedback}, we can guide the search of $\lambda$ based on value of SNIPS denominator $S$. 
It was shown in \cite{deep-learning-logged-bandit-feedback} that the value of $S$ increases monotonically, if $\lambda$ is increased. 
Secondly, it is straightforward to note that expectation of $S$ is 1. 
This implies that, with increasing number of bandit feedback, the optimal value for $\lambda$ should be selected such that its corresponding $S$ value concentrates around 1. 
In our experiments, we first select some random $\lambda \in [0, 1]$ and train the model for two epochs with this $\lambda$. We then calculate $S$ for the trained model; if $S$ is greater than 1, we decrease $\lambda$ by 10\%, otherwise we increase it by 10\%. The final value of $\lambda$ is decided based on best performance on validation set.

In Fig.~\ref{fig:Sandlambda}, we plot the values of denominator $S$ on order logs (training set) of \textit{Mercateo dataset} for different values of hyperparameter $\lambda$. On the figure below, Fig.~\ref{fig:Metricsandlambda}, we also plot performance on orders test set, in terms of MAP and NDCG@5 scores, of different rankers for these values of hyperparameter $\lambda$.
It is to be noted that the values of SNIPS denominator $S$ monotonicaly increase with increasing $\lambda$. 
The MAP and NDCG@5 reach its highest value for $\lambda = 0.4$, but decrease only slightly with increasing values of $\lambda$. 
Furthermore, it can also be seen from these two figures that the $\lambda$ values with good performance on test set have corresponding SNIPS denominator values close to 1.

\end{document}

%% file: plots/average_rank.tikz
\resizebox{70mm}{60mm}{
\begin{tikzpicture}

\begin{axis}[
axis background/.style={fill=white!89.80392156862746!black},
axis line style={white},
legend cell align={left},
legend entries={{CRM},{CE}},
legend style={draw=white!80.0!black, fill=white!89.80392156862746!black},
tick align=outside,
tick pos=left,
x grid style={white},
xlabel={number of bandit feedback [millions]},
xmajorgrids,
xmin=-228999.509856631, xmax=70154355.0985663,
xtick={0,10000000,20000000,30000000,40000000,50000000,60000000,70000000},
scaled x ticks = false,
x tick label style={/pgf/number format/fixed},
xticklabels={0,1,2,3,4,5,6,7},
y grid style={white},
ylabel={Average Rank},
ymajorgrids,
ymin=1.01469246666161, ymax=15.1146037690176
]
\addlegendimage{no markers, green!50.19607843137255!black}
\addlegendimage{no markers, red}
\addplot [only marks, draw=green!50.19607843137255!black, fill=green!50.19607843137255!black, colormap/viridis]
table [row sep=\\]{%
x                      y\\ 
+6.143470000000000e+05 +1.445990180032733e+01\\
+1.228650000000000e+06 +1.434206219312602e+01\\
+2.457157000000000e+06 +1.262258592101e+01 \\
+6.757949000000000e+06 +1.162288379701e+01 \\
+7.372227000000000e+06 +9.5227587152210e+00 \\
+9.215182000000000e+06 +7.321787152210e+00 \\
+1.036339900000000e+07 +5.698854337152210e+00\\
+2.625605300000000e+07  +2.955973813420622e+00\\
+3.600510500000000e+07  +2.728314238952537e+00\\
+4.391110700000000e+07 +2.66612111229623e+00\\
+4.935960700000000e+07 +2.486088379705401e+00\\
+5.243145400000000e+07  +2.088379705400982e+00\\
+6.686143000000000e+07 +1.669394435351882e+00\\
};
\addplot [semithick, green!50.19607843137255!black]
table [row sep=\\]{%
614347	14.4599018003273 \\
1228650	14.342062193126 \\
2457157	12.6225859247136 \\
6757949	11.6225859247136 \\
7372227	9.5225859247136 \\
9215182	7.3225859247136 \\
10363399	5.69885433715221 \\
26256053	2.95597381342062 \\
36005105	2.72831423895254 \\
43911107	2.66612111292962 \\
49359607	2.4860883797054 \\
52431454	2.08837970540098 \\
66861430	1.66939443535188 \\
};	
\addplot [semithick, red]
table [row sep=\\]{%
614347	10.93 \\
1228650	10.93 \\
2457157	10.93 \\
6757949	10.93 \\
7372227	10.93 \\
9215182	10.93 \\
10363399	10.93 \\
26256053	10.93 \\
36005105	10.93 \\
43911107	10.93 \\
49359607	10.93 \\
52431454	10.93 \\
66861430	10.93 \\
};	
\end{axis}

\end{tikzpicture}}

%% file: plots/dcg_1.tikz
\resizebox{70mm}{60mm}{
\begin{tikzpicture}

\begin{axis}[
axis background/.style={fill=white!89.80392156862746!black},
axis line style={white},
legend cell align={left},
legend entries={{CRM},{CE}},
legend style={at={(0.97,0.5)}, anchor=east, draw=white!80.0!black, fill=white!89.80392156862746!black},
tick align=outside,
tick pos=left,
x grid style={white},
scaled x ticks = false,
x tick label style={/pgf/number format/fixed},
xlabel={number of bandit feedback [millions]},
xmajorgrids,
xmin=-3105001.30985663, xmax=70154355.0985663,
xtick={0,10000000,20000000,30000000,40000000,50000000,60000000,70000000},
xticklabels={0,1,2,3,4,5,6,7},
y grid style={white},
ylabel={DCG},
ymajorgrids,
ymin=1.41512882600486, ymax=4.03584719930722
]
\addlegendimage{no markers, green!50.19607843137255!black}
\addlegendimage{no markers, red}
\addplot [only marks, draw=green!50.19607843137255!black, fill=green!50.19607843137255!black, colormap/viridis]
table [row sep=\\]{%
x                      y\\ 
+6.143470000000000e+05 +1.548049297919966e+00\\
+1.228650000000000e+06 +1.570563621587533e+00\\
+2.457157000000000e+06 +1.631951594522356e+00\\
+6.757949000000000e+06 +1.7300808412952e+00\\
+7.372227000000000e+06 +2.365028607558621e+00\\
+9.215182000000000e+06 +2.595898525857729e+00\\
+1.036339900000000e+07 +2.864644906979341e+00\\
+2.625605300000000e+07 +3.242699727791468e+00\\
+3.600510500000000e+07 +3.750275011921466e+00\\
+4.391110700000000e+07 +3.788561830650053e+00\\
+4.935960700000000e+07 +3.858738128886996e+00\\
+5.243145400000000e+07 +3.879822796257675e+00\\
+6.686143000000000e+07 +3.902926727392111e+00\\
};
\addplot [semithick, green!50.19607843137255!black]
table [row sep=\\]{%
614347	1.54804929791997 \\
1228650	1.57056362158753 \\
2457157	1.63195159452236 \\
6757949	1.730080841295 \\
7372227	2.36502860755862 \\
9215182	2.59589852585773 \\
10363399	2.86464490697934 \\
26256053	3.24269972779147 \\
36005105	3.75027501192147 \\
43911107	3.78856183065005 \\
49359607	3.858738128887 \\
52431454	3.87982279625768 \\
66861430	3.90292672739211 \\
};
\addplot [semithick, red]
table [row sep=\\]{%
614347	1.78 \\
1228650	1.78 \\
2457157	1.78 \\
6757949	1.78 \\
7372227	1.78 \\
9215182	1.78 \\
10363399	1.78 \\
26256053	1.78 \\
36005105	1.78 \\
43911107	1.78 \\
49359607	1.78 \\
52431454	1.78 \\
66861430	1.78 \\
};
\end{axis}

\end{tikzpicture}}

%% file: plots/snips_denominator.tikz
\begin{tikzpicture}

\definecolor{color0}{rgb}{0.886274509803922,0.290196078431373,0.2}

\begin{axis}[
axis background/.style={fill=white!89.80392156862746!black},
axis line style={white},
tick align=outside,
tick pos=left,
x grid style={white},
xlabel={hyperparameter $\lambda$},
xmajorgrids,
xmin=0.152643369175627, xmax=0.997356630824373,
y grid style={white},
ylabel={SNIPS denominator},
ymajorgrids,
ymin=-0.0681793516415011, ymax=1.2050090076415
]
\addplot [only marks, draw=color0, fill=color0, colormap/viridis]
table [row sep=\\]{%
x                      y\\ 
+2.000000000000000e-01 +3.489756000000000e-03\\ 
+3.000000000000000e-01 +5.196917000000000e-02\\ 
+4.000000000000000e-01 +7.862210000000000e-01\\
+7.000000000000000e-01 +9.856925500000000e-01\\ 
+9.000000000000000e-01 +1.098162900000000e+00\\ 
+9.500000000000000e-01 +1.133339900000000e+00\\ 
};
\addplot [semithick, color0, forget plot]
table [row sep=\\]{%
0.2	0.003489756 \\
0.3	0.05196917 \\
0.4	7.862210000e-01\\
0.7	+9.85692550e-01 \\
0.9	1.0981629 \\
0.95	1.1333399 \\
};
\end{axis}

\end{tikzpicture}

%% file: plots/lagrange_map_ndcg.tikz
\begin{tikzpicture}

\begin{axis}[
axis background/.style={fill=white!89.80392156862746!black},
axis line style={white},
legend cell align={left},
legend entries={{CRM MAP},{CRM NDCG@5}},
legend style={at={(0.97,0.03)}, anchor=south east, draw=white!80.0!black, fill=white!89.80392156862746!black},
tick align=outside,
tick pos=left,
x grid style={white},
xlabel={hyperparameter $\lambda$},
xmajorgrids,
xmin=0.1625, xmax=0.9875,
y grid style={white},
ylabel={MAP, NDCG},
ymajorgrids,
ymin=0.06778, ymax=0.33002
]
\addlegendimage{no markers, green!50.19607843137255!black}
\addlegendimage{no markers, red}
\addplot [semithick, green!50.19607843137255!black]
table [row sep=\\]{%
0.2	0.1431 \\
0.3	0.1574 \\
0.4	0.2941 \\
0.5	0.3147 \\
0.7	0.29 \\
0.9	0.281 \\
0.95	0.2657 \\
};
\addplot [semithick, red]
table [row sep=\\]{%
0.2	0.0797 \\
0.3	0.1275 \\
0.4	0.3181 \\
0.5	0.293 \\
0.7	0.2792 \\
0.9	0.2801 \\
0.95	0.2609 \\
};
\end{axis}

\end{tikzpicture}